\documentclass[11pt]{SciPost_better_arXiv}

\hypersetup{
    colorlinks,
    linkcolor={red!50!black},
    citecolor={blue!50!black},
    urlcolor={blue!80!black}
}

\usepackage[bitstream-charter]{mathdesign}
\usepackage{booktabs}
\usepackage[export]{adjustbox}
\usepackage{subcaption}
\usepackage{amsmath}

\DeclareSymbolFont{usualmathcal}{OMS}{cmsy}{m}{n}
\DeclareSymbolFontAlphabet{\mathcal}{usualmathcal}

\fancypagestyle{SPstyle}{
\fancyhf{}
\lhead{\colorbox{scipostblue}{\bf \color{white} ~SciPost Physics Community Reports }}
\rhead{{\bf \color{scipostdeepblue} ~Submission }}

\fancyfoot[C]{\textbf{\thepage}}
}

\RequirePackage{heppennames2}
\RequirePackage{lhchiggs}

\usepackage{graphicx}
\usepackage{tabularx}
\usepackage{subcaption}
\usepackage{placeins}
\usepackage{multirow}
\usepackage{graphbox}
\usepackage{amsmath}
\usepackage{mathtools}
\usepackage{mathrsfs}
\usepackage{hyperref}
\newcommand{\GeV}{\ensuremath{\,\mathrm{GeV}}}

\usepackage[dvipsnames,table]{xcolor}

\date{\today}

\begin{document}
\pagestyle{SPstyle}
\begin{flushright}\footnotesize{LHCHWG-2026-001, CERN-TH-2026-007, FR-PHENO-26-02, OUTP-26-01P, COMETA-2026-02}\end{flushright}
\begin{center}{\Large \textbf{\color{scipostdeepblue}{
Electroweak Higgs boson pair production: \\
Updated inclusive cross sections
}}}\end{center}

\begin{center}\textbf{
Ramona~Gr\"ober\textsuperscript{1,2},
Alexander~Karlberg\textsuperscript{3,4},
Mathieu~Pellen\textsuperscript{5},
Gioia~Sacchi\textsuperscript{1},
Michael~Spira\textsuperscript{6}
}\end{center}

\begin{center}
{\bf 1} Dipartimento di Fisica e Astronomia `G.Galilei', Universit\`a di Padova, Via F Marzolo, 35131 Padova, Italy
\\
{\bf 2} INFN, Sezione di Padova, Via F Marzolo, 35131 Padova, Italy
\\
{\bf 3} Rudolf Peierls Centre for Theoretical Physics, Clarendon Laboratory, Parks Road, OX1 3PU, Oxford, UK
\\
{\bf 4} CERN, Theoretical Physics Department, 1211, Geneva 23, Switzerland
\\
{\bf 5} Albert-Ludwigs-Universität Freiburg, Physikalisches Institut, Hermann-Herder-Str. 3, 79104
Freiburg, Germany
\\
{\bf 6} PSI Center for Neutron and Muon Sciences, 5232 Villigen PSI, Switzerland

\href{mailto:ramona.groeber@pd.infn.it}{\small ramona.groeber@pd.infn.it},
\href{mailto:alexander.karlberg@cern.ch}{\small alexander.karlberg@cern.ch},
\href{mailto:mathieu.pellen@physik.uni-freiburg.de}{\small mathieu.pellen@physik.uni-freiburg.de},
\href{mailto:gioia.sacchi@studenti.unipd.it}{\small gioia.sacchi@studenti.unipd.it},
\href{mailto:Michael.Spira@psi.ch}{\small
michael.spira@psi.ch}
\end{center}

\section*{\color{scipostdeepblue}{Abstract}}
\textbf{\boldmath{%
We present updated inclusive cross sections for electroweak Higgs boson pair production for energies of relevance to the LHC and High-Luminosity phase of the LHC. The cross sections are presented at N$^3$LO QCD+NLO EW for vector-boson fusion and NNLO QCD for associate production with a vector boson. We compute the cross sections using the most up-to-date theory inputs, both in the Standard Model and for a few anomalous values of the trilinear Higgs self-coupling. 
}}

\vspace{\baselineskip}

\noindent\textcolor{white!90!black}{%
\fbox{\parbox{0.975\linewidth}{%
\textcolor{white!40!black}{\begin{tabular}{lr}%
  \begin{minipage}{0.6\textwidth}%
    {\small Copyright attribution to authors. \newline
    This work is a submission to SciPost Phys. Comm. Rep. \newline
    License information to appear upon publication. \newline
    Publication information to appear upon publication.}
  \end{minipage} & \begin{minipage}{0.4\textwidth}
    {\small Received Date \newline Accepted Date \newline Published Date}%
  \end{minipage}
\end{tabular}}
}}
}


\vspace{10pt}
\noindent\rule{\textwidth}{1pt}
\tableofcontents
\noindent\rule{\textwidth}{1pt}
\vspace{10pt}

\section{Introduction}
With the discovery of the Higgs boson by the ATLAS~\cite{ATLAS:2012yve} and CMS~\cite{CMS:2012qbp} Collaborations at the CERN Large Hadron Collider (LHC) in 2012, we are now preparing for the upcoming high luminosity phase of the LHC (HL-LHC)~\cite{Apollinari:2017lan}, with a particular view to unravel the Higgs sector. One of the main objectives of the HL-LHC will be to precisely determine the trilinear Higgs self-coupling~\cite{Cepeda:2019klc}. Although gluon-gluon-fusion Higgs boson pair production (ggF) dominates the cross section, other production channels such as associated production with a vector boson ($Vhh$) and vector-boson fusion (VBF) ($hhjj$) can give complementary information from the point of view of measuring anomalous couplings. 

In this note, we present updated theoretical predictions for $Vhh$ and VBF Higgs boson pair production in proton-proton collisions at the (HL)-LHC. The measurements of both processes give access to various effective couplings as for instance the coupling $\kappa_{2V}$ of two Higgs bosons to two vector bosons. It is interesting to note that, Higgs boson pair production in association with a vector boson, $pp\to Vhh$, allows to constrain $\kappa_{2V}$ for $W$ and $Z$ bosons independently, whereas in VBF Higgs boson pair production this is not possible, but on the other hand it provides larger cross sections. It is therefore of importance to measure both electroweak (EW) Higgs boson pair production modes.

The production of two Higgs bosons via VBF is the second largest production mechanism in terms of cross section.
Therefore the sensitivity to Higgs boson pair production via VBF, possibly in association with the ggF mechanism, has already been studied experimentally by both the ATLAS~\cite{ATLAS:2022jtk,ATLAS:2023gzn,ATLAS:2023qzf,ATLAS:2024lsk,ATLAS:2024ish,ATLAS:2024pov,ATLAS:2024lhu,ATLAS:2025hhd} and CMS~\cite{CMS:2017yfv,CMS:2020tkr,CMS:2022cpr,CMS:2022ngs,CMS:2022gjd,CMS:2024xus,CMS:2024ymd,CMS:2024fkb,CMS:2024rgy} Collaborations with the current LHC data sets.
Sensitivity studies for the high luminosity phase of the LHC have also been carried out~\cite{ATLAS:2024voi,ATLAS:2025bsu,ATLAS:2025cwr,ATLAS:2025wdq, ATLAS:2025eii}. While sensitivity to the SM cross section cannot be reached even at the HL-LHC, those sensitivity studies are interpreted in terms of potential bounds on anomalous couplings. 
\par
Although $Vhh$ production ranks only fourth in terms of the Higgs boson pair production cross sections, both ATLAS and CMS have initiated dedicated searches, currently setting upper limits of
$187\times \sigma_{SM}$
\cite{ATLAS:2022fpx} and $294\times \sigma_{SM}$
\cite{CMS:2024fkb} at 95\% confidence level (C.L.).

We note that since we present cross sections updates based on existing theoretical calculations, this note is rather short. We therefore refer the reader to the original references for details on the various calculations.

\section{Calculational setup}

In this section, we reproduce the general setup of the calculations presented here, following the guidelines of the Higgs cross section working group.\footnote{See \url{https://twiki.cern.ch/twiki/bin/view/LHCPhysics/LHCHWG136TeVxsec}.}

The parton distribution function (PDF) \texttt{PDF4LHC21\_40}~\cite{Cridge:2021qjj,PDF4LHCWorkingGroup:2022cjn} with $\alphas(M_{\PZ})= 0.118$ is utilised through the use of \textsc{LHAPDF}~\cite{Buckley:2014ana}.
The masses and widths used are:
\begin{align}
                  m_t   &=  172.5\GeV,       & \quad \quad \quad \Mb &= 0 \GeV,  \nonumber \\
                \MZ &=  91.1876\GeV,      & \quad \quad \quad \GZ &= 2.4952\GeV,  \nonumber \\
                \MW &=  80.379\GeV,       & \GW &= 2.085\GeV.
\end{align}
The Higgs boson width is set to zero.
The EW coupling is fixed in the $\GF$ scheme \cite{Denner:2000bj,Dittmaier:2001ay} upon using:
\begin{align}
  \alpha = \frac{\sqrt{2}}{\pi} \GF \MW^2 \left( 1 - \frac{\MW^2}{\MZ^2} \right)  \qquad \mathrm{and}  
  \qquad   \GF    = 1.16638\times 10^{-5}~\mbox{GeV}^{-2}\;.
\end{align}
The numerical value thus obtained reads $\alpha = 0.75652103079904 \times 10^{-2}$.

\section{Updated cross section values for VBF $hh$ production}
In this section we present cross sections for VBF Higgs boson pair production. We define VBF production in the usual sense, where at leading order (LO) $s$-channel contributions of the type $pp\to hh (V\to jj)$, where $V$ refers to either a $W$ or $Z$ gauge boson, have been omitted, meaning that only $t$ and $u$ contributions are retained.
In addition, the square of the latter two topologies is taken separately and interferences between them are omitted.\footnote{More discussion on the definition and the validity of the VBF approximations can be found for example in Ref.~\cite{Barone:2025jey}.} At higher orders we work in the so-called factorised approximation, where the two VBF quark lines are assumed to be completely independent. Next-to LO (NLO) QCD predictions in this approximation were computed in~\cite{Figy:2008zd,Baglio:2012np,Frederix:2014hta}.
Nowadays, state-of-the-art predictions at fixed order comprise N$^3$LO QCD corrections for inclusive production~\cite{Ling:2014sne,Dreyer:2018qbw} and NNLO QCD corrections differentially~\cite{Dreyer:2018rfu}.
In addition, in the last few years non-factorisable QCD corrections~\cite{Dreyer:2020urf} have been computed along with NLO EW corrections~\cite{Dreyer:2020xaj}. The latter were combined with NNLO QCD corrections in the same reference.
When including parton-shower (PS) corrections, the best accuracy to day is NLO QCD+PS~\cite{Frederix:2014hta,Jager:2025isz}, and in general the parton shower has a moderate effect on the distributions. It is worth noting that these computations are publicly available.

The inclusive predictions provided in Table~\ref{tab:hhjj} are at N$^3$LO QCD+NLO EW accuracy and were obtained with \textsc{proVBFHH} v2.1.0~\cite{Cacciari:2015jma,Dreyer:2016oyx,Dreyer:2018rfu,Dreyer:2018qbw,Dreyer:2020xaj} and \textsc{Recola+MoCaNLO}~\cite{Actis:2016mpe,Denner:2026phn}, respectively. \textsc{proVBFHH} is based on \textsc{Hoppet}~\cite{Salam:2008qg,Karlberg:2025hxk} and makes use of the parametrised two- and three-loop deep inelastic scattering coefficient functions from \cite{vanNeerven:1999ca,vanNeerven:2000uj,Vermaseren:2005qc,Moch:2004xu,Moch:2008fj,Davies:2016ruz} for the cross sections presented here.

\begin{table}
\renewcommand{\arraystretch}{1.5}\small
\centering
\begin{tabular}{c| c |c }
$\sqrt{s}$ [TeV] & $m_h$ [GeV] & $\sigma $ [fb] SM \\ 
\toprule
     & 125    & $1.687^{+0.05\%}_{-0.04\%} \pm 2.7\%$ \\
13   & 125.09 & $1.684^{+0.05\%}_{-0.04\%} \pm 2.7\%$  \\
     & 125.38 & $1.676^{+0.05\%}_{-0.04\%} \pm 2.8\%$  \\ 
\midrule
     & 125    & $1.874^{+0.05\%}_{-0.03\%} \pm 2.7\%$ \\
13.6 & 125.09 & $1.870^{+0.05\%}_{-0.03\%} \pm 2.7\%$ \\
     & 125.38 & $1.861^{+0.05\%}_{-0.03\%} \pm 2.7\%$ \\
\midrule
     & 125    & $2.005^{+0.05\%}_{-0.03\%} \pm 2.7\%$ \\
14   & 125.09 & $2.001^{+0.05\%}_{-0.03\%} \pm 2.7\%$ \\
     & 125.38 & $1.992^{+0.05\%}_{-0.03\%} \pm 2.7\%$ \\
\end{tabular}
\caption{Cross section values for VBF $hh$ production including scale uncertainties and PDF uncertainties in percent at N$^3$LO QCD+NLO EW in the SM. Note that no EW uncertainties are included, but that they are assumed to be fully contained in the PDF uncertainty which dominates. 
Upon using these numbers, \cite{Dreyer:2018qbw} and \cite{Dreyer:2020xaj} should be cited as a minimum.
\label{tab:hhjj} }
\end{table}

Cross section values with anomalous trilinear Higgs self-coupling $\lambda_{hhh}$ can also be obtained by \textsc{proVBFHH} but without NLO EW corrections. In Table~\ref{tab:hhjj-anom} we report cross section values with $\kappa_\lambda=\lambda_{hhh}/\lambda_{hhh}^{\mathrm{SM}}=\{0,1,2,3\}$, where $\lambda_{hhh}^{\mathrm{SM}}$ is the SM value of the trilinear Higgs self-coupling. For more detailed studies of the impact of anomalous couplings in the framework of the Higgs Effective Field Theory we refer to \cite{Englert:2023uug,Jager:2025isz,Braun:2025hvr,Braun:2942975} and to \cite{Dedes:2025oda} for a study of VBF Higgs boson pair production in the Standard Model Effective Field Theory up to dimension-eight.

The renormalisation and factorisation scale used for VBF Higgs boson pair production is defined as
\begin{align}
    \mu = \sqrt{\frac{m_h}{2} \sqrt{\left( \frac{m_h} {2}\right)^2 + p^2_{\rm T, hh}}},
\end{align}
where $p_{\rm T, hh}$ is the transverse momentum of the di-Higgs boson system.
It is worth emphasising that the predictions are done consistently within the VBF approximation.
It also means that, in contrast to Ref.~\cite{Dreyer:2020xaj} where full NLO QCD corrections were included, only VBF-approximate predictions are utilised and combined in a multiplicative way as
\begin{align}
    \sigma^{\rm VBF}_{\rm N3LO\; QCD \times NLO\; EW} = \sigma^{\rm VBF}_{\rm N3LO\; QCD} \left(1+\frac{\delta^{\rm VBF}_{\rm NLO\; EW}}{\sigma^{\rm vbf}_{\rm LO}} \right) ,
\end{align}
with $\sigma^{\rm VBF}_{\rm NLO\; EW} = \sigma^{\rm vbf}_{\rm LO} + \delta^{\rm VBF}_{\rm NLO\; EW}$.
Also, the EW corrections computed here do not include photon-induced contributions.
The reasons for this choice are manifold.
First, the recommended PDF set does not include a photon PDF meaning that using another PDF would be inconsistent.
Secondly, photon-induced contributions have been found to be small~\cite{Barone:2025jey} for VBF production in the  $\rm \overline{MS}$ scheme (well below $1\%$).
Finally, on the technical side, demanding the VBF approximation for photon-induced contributions would require generating specific approximated matrix elements not available in \textsc{Recola}.

\begin{table}
\renewcommand{\arraystretch}{1.5}
\small
\centerline{
\begin{tabular}{c| c |c | c | c| c }
$\sqrt{s}$ [TeV] & $m_h$ [GeV] & $\sigma $ [fb] SM & $\sigma $ [fb] $\kappa_{\lambda}=0$ & $\sigma $ [fb] $\kappa_{\lambda}=2$  & $\sigma $ [fb] $\kappa_{\lambda}=3$  \\ \toprule
     & 125    & $1.742^{+0.04\%}_{-0.03\%} \pm 2.7\%$ & $4.543^{+0.02\%}_{-0.01\%} \pm 2.6\%$ & $1.443^{+0.06\%}_{-0.04\%} \pm 2.7\%$ & $3.647^{+0.04\%}_{-0.00\%} \pm 2.5 \%$ \\
13   & 125.09 & $1.739^{+0.04\%}_{-0.03\%} \pm 2.7\%$ & $4.540^{+0.02\%}_{-0.01\%} \pm 2.6\%$ & $1.442^{+0.06\%}_{-0.04\%} \pm 2.7\%$ & $3.648^{+0.04\%}_{-0.00\%} \pm 2.5 \%$ \\
     & 125.38 & $1.731^{+0.04\%}_{-0.03\%} \pm 2.7\%$ & $4.528^{+0.02\%}_{-0.01\%} \pm 2.6\%$ & $1.440^{+0.06\%}_{-0.04\%} \pm 2.7\%$ & $3.652^{+0.04\%}_{-0.00\%} \pm 2.5 \%$ \\ 
\midrule
     & 125    & $1.938^{+0.03\%}_{-0.03\%} \pm 2.7\%$ & $5.019^{+0.02\%}_{-0.01\%} \pm 2.5\%$ & $1.606^{+0.05\%}_{-0.04\%} \pm 2.7\%$ & $4.022^{+0.04\%}_{-0.00\%} \pm 2.5 \%$ \\
13.6 & 125.09 & $1.936^{+0.03\%}_{-0.03\%} \pm 2.7\%$ & $5.015^{+0.02\%}_{-0.01\%} \pm 2.5\%$ & $1.605^{+0.05\%}_{-0.04\%} \pm 2.7\%$ & $4.023^{+0.04\%}_{-0.00\%} \pm 2.5 \%$ \\
     & 125.38 & $1.927^{+0.03\%}_{-0.03\%} \pm 2.7\%$ & $5.002^{+0.02\%}_{-0.01\%} \pm 2.5\%$ & $1.602^{+0.05\%}_{-0.04\%} \pm 2.7\%$ & $4.028^{+0.04\%}_{-0.00\%} \pm 2.5 \%$ \\
\midrule
     & 125    & $2.075^{+0.03\%}_{-0.02\%} \pm 2.7\%$ & $5.347^{+0.02\%}_{-0.01\%} \pm 2.5\%$ & $1.719^{+0.05\%}_{-0.03\%} \pm 2.7\%$ & $4.281^{+0.04\%}_{-0.00\%} \pm 2.5 \%$ \\
14   & 125.09 & $2.072^{+0.03\%}_{-0.02\%} \pm 2.7\%$ & $5.342^{+0.02\%}_{-0.01\%} \pm 2.5\%$ & $1.719^{+0.05\%}_{-0.03\%} \pm 2.7\%$ & $4.282^{+0.04\%}_{-0.00\%} \pm 2.5 \%$ \\
     & 125.38 & $2.022^{+0.03\%}_{-0.02\%} \pm 2.7\%$ & $5.329^{+0.02\%}_{-0.01\%} \pm 2.5\%$ & $1.715^{+0.05\%}_{-0.03\%} \pm 2.7\%$ & $4.287^{+0.04\%}_{-0.00\%} \pm 2.5 \%$ \\
\end{tabular}}
\caption{Cross section values for VBF Higgs boson pair production including scale uncertainties and PDF uncertainties in percent at N$^3$LO QCD in the SM ($\kappa_\lambda = 1$) and for three anomalous values of the trilinear coupling $\kappa_\lambda=\{0,2,3\}$. Note that no EW corrections are included. \label{tab:hhjj-anom} }
\end{table}

\section{Updated cross section values for associated production with a vector boson}
We updated the current best SM predictions to a modern PDF set \cite{PDF4LHCWorkingGroup:2022cjn}, provide with respect to \cite{LHCHiggsCrossSectionWorkingGroup:2016ypw} numbers for a centre-of-mass energy of $\sqrt{s}= 13.6$ TeV and for various values of $\kappa_{\lambda}$. For this purpose we use the inclusive NNLO QCD corrected cross sections provided in \cite{Baglio:2012np}, which should be cited if using the numbers of this report. We use as central scale $\mu_R=\mu_F=M_{Vhh}$, the invariant mass of the $Vhh$ system. Fully differential NNLO QCD corrected cross sections were provided in \cite{Li:2016nrr, Li:2017lbf}. 

\subsection{$W^{\pm}hh$}
The process can be regarded as Drell-Yan production
$pp \to V^*$ followed by the splitting process $V^* \to V hh$. 
The NNLO QCD corrections were hence  obtained adjusting the known Drell-Yan NNLO QCD corrections \cite{Hamberg:1990np, Harlander:2002wh}, in analogy to $Vh$ production \cite{Brein:2003wg}.\footnote{We note that in the meanwhile also N$^3$LO corrections to Drell-Yan production are known \cite{Baglio:2022wzu}.} In Table \ref{tab:Wphh} and \ref{tab:Wmhh} we provide updated cross section values for $W^+ hh$ and $W^- hh$ production respectively, using the \texttt{PDF4LHC21} PDF set \cite{PDF4LHCWorkingGroup:2022cjn}. We observe a larger scale uncertainty for $W^-hh$ compared to $W^+hh$.
Since the two processes receive different relative contributions from the $q\bar{q}$
  and $qg$ initial states, which nearly cancel,\footnote{The $gg$ initial state gives a very small contribution only.} it is not surprising that the residual scale variation differs.
These differences originate from the distinct parton–luminosity weights in the proton PDFs. 
\begin{table}
\renewcommand{\arraystretch}{1.5}\small
\centerline{
\begin{tabular}{c| c |c|c| c| c }
$\sqrt{s}$ [TeV] & $m_h$ [GeV] & $\sigma $ [fb] SM & $\sigma $ [fb] $\kappa_{\lambda}=0$ & $\sigma $ [fb] $\kappa_{\lambda}=2$ & $\sigma $ [fb] $\kappa_{\lambda}=3$\\ \hline
13 & 125 & $0.334^{+ 0.34\%}_{-0.41 \%} \pm 2.3 \%$ & $0.192^{+ 0.31\%}_{-0.42 \%} \pm 2.4 \%$& $0.549^{+ 0.37\%}_{-0.41 \%} \pm 2.3 \%$  & $0.837^{+ 0.37\%}_{-0.41 \%} \pm 2.3 \%$  \\
13 & 125.09 & $0.333^{ +0.34\%}_{-0.41 \%} \pm 2.3\%$ & $0.192^{+ 0.31\%}_{-0.42 \%} \pm 2.4 \%$& $0.548^{+ 0.37\%}_{-0.41 \%} \pm 2.3 \%$  & $0.835^{+ 0.37\%}_{-0.41 \%} \pm 2.3 \%$  \\
13 & 125.38 & $0.331 ^{+0.34 \%}_{-0.41 \%} \pm 2.3 \%$ & $0.190^{+ 0.31\%}_{-0.42 \%} \pm 2.4 \%$& $0.544^{+ 0.37\%}_{-0.41 \%} \pm 2.3 \%$  & $0.831^{+ 0.37\%}_{-0.41 \%} \pm 2.3 \%$  \\ \midrule
13.6 & 125 & $0.358 ^{+0.36 \%}_{-0.40 \%}  \pm 2.3 \%$& $0.206 ^{+0.33 \%}_{-0.41 \%}  \pm 2.3 \% $ & $ 0.588 ^{+0.38 \%}_{-0.40 \%}  \pm 2.2 \%  $ &  $ 0.896 ^{+0.39 \%}_{-0.40 \%}  \pm 2.2 \%   $\\
13.6 & 125.09 & $ 0.357^{+0.36\%}_{-0.40\%}  \pm 2.3 \% $ & $0.206 ^{+0.33 \%}_{-0.41 \%}  \pm 2.3 \% $ & $ 0.587 ^{+0.38 \%}_{-0.40 \%}  \pm 2.2 \%  $ &  $ 0.895 ^{+0.39 \%}_{-0.40 \%}  \pm 2.2 \%   $ \\
13.6 & 125.38 & $0.355^{+0.36 \%}_{-0.40 \%}  \pm 2.3 \% $ & $0.204 ^{+0.33 \%}_{-0.41 \%}  \pm 2.3 \% $ & $ 0.583 ^{+0.38 \%}_{-0.40 \%}  \pm 2.2 \%  $ &  $ 0.890 ^{+0.39 \%}_{-0.40 \%}  \pm 2.2 \%   $\\ \midrule
14 & 125 &  $0.374^{+0.37\%}_{-0.40 \%} \pm 2.2 \% $ & $0.216 ^{+0.33 \%}_{-0.40 \%}  \pm 2.3 \% $ & $ 0.614 ^{+0.39 \%}_{-0.40 \%}  \pm 2.2 \%  $ &  $ 0.936 ^{+0.40 \%}_{-0.41 \%}  \pm 2.2 \%   $  \\
14 & 125.09 & $0.373^{+0.37 \%}_{-0.40 \%} \pm 2.2 \% $ &$0.216 ^{+0.33 \%}_{-0.40 \%}  \pm 2.3 \% $ & $ 0.613 ^{+0.39 \%}_{-0.40 \%}  \pm 2.2 \%  $ &  $ 0.935 ^{+0.40 \%}_{-0.41 \%}  \pm 2.2 \%   $  \\
14 & 125.38 &  $ 0.371^{+0.37 \%}_{-0.40\%} \pm 2.2 \% $ & $0.214 ^{+0.33 \%}_{-0.40 \%}  \pm 2.3 \% $ & $ 0.610 ^{+0.39 \%}_{-0.40 \%}  \pm 2.2 \%  $ &  $ 0.930 ^{+0.40 \%}_{-0.41 \%}  \pm 2.2 \%   $\\
\end{tabular}}
\caption{Cross section values for $W^+hh$ production including scale uncertainties and PDF uncertainties in percent in this order at NNLO QCD in the SM ($\kappa_\lambda = 1$) and for three anomalous values of the trilinear coupling $\kappa_\lambda=\{0,2,3\}$. \label{tab:Wphh}}
\end{table}

\begin{table}
\renewcommand{\arraystretch}{1.5}
\small
\centerline{
\begin{tabular}{c| c |c | c | c | c   }
$\sqrt{s}$ [TeV] & $m_h$ [GeV] & $\sigma $ [fb] SM & $\sigma $ [fb] $\kappa_{\lambda}=0$ & $\sigma $ [fb] $\kappa_{\lambda}=2$ & $\sigma $ [fb] $\kappa_{\lambda}=3$ \\ \hline
13 & 125 & $0.173^{+ 1.3\%}_{-1.3 \%} \pm 2.5 \%$ & $0.098^{+ 1.3\%}_{-1.3 \%} \pm 2.6 \%$ &  $0.286^{+ 1.3\%}_{-1.3 \%} \pm 2.5 \%$ & $0.437^{+ 1.3\%}_{-1.4 \%} \pm 2.5 \% $\\
13 & 125.09 & $0.172^{ +1.3\%}_{-1.3 \%} \pm 2.5\%$ & $0.098^{+ 1.3\%}_{-1.3 \%} \pm 2.6 \%$ &  $0.285^{+ 1.3\%}_{-1.3 \%} \pm 2.5 \%$ & $0.436^{+ 1.3\%}_{-1.4 \%} \pm 2.5 \% $ \\
13 & 125.38 & $0.171 ^{+1.3 \%}_{-1.3 \%} \pm 2.5 \%$ & $0.097^{+ 1.3\%}_{-1.3 \%} \pm 2.6 \%$ &  $0.283^{+ 1.3\%}_{-1.3 \%} \pm 2.5 \%$ & $0.433^{+ 1.3\%}_{-1.4 \%} \pm 2.5 \% $\\ \midrule
13.6 & 125 & $0.187 ^{+1.3 \%}_{-1.4 \%}  \pm 2.5 \%$ & $0.106^{+ 1.3\%}_{-1.3 \%} \pm 2.6 \%$ &  $0.309^{+ 1.3\%}_{-1.4 \%} \pm 2.4 \%$ & $0.472^{+ 1.3\%}_{-1.4 \%} \pm 2.4 \% $\\
13.6 & 125.09 & $ 0.187^{+1.3\%}_{-1.4\%}  \pm 2.5 \% $  & $0.106^{+ 1.3\%}_{-1.3 \%} \pm 2.6 \%$ &  $0.309^{+ 1.3\%}_{-1.4 \%} \pm 2.4 \%$ & $0.472^{+ 1.3\%}_{-1.4 \%} \pm 2.4 \% $ \\
13.6 & 125.38 & $0.185^{+1.3 \%}_{-1.4 \%}  \pm 2.5 \% $ & $0.105^{+ 1.3\%}_{-1.3 \%} \pm 2.6 \%$ &  $0.307^{+ 1.3\%}_{-1.4 \%} \pm 2.4 \%$ & $0.469^{+ 1.3\%}_{-1.4 \%} \pm 2.4 \% $\\ \midrule
14 & 125 &  $0.197^{+1.3\%}_{-1.4 \%} \pm 2.4 \% $ & $0.112^{+ 1.3\%}_{-1.3 \%} \pm 2.5 \%$ &  $0.325^{+ 1.3\%}_{-1.4 \%} \pm 2.4 \%$ & $0.497^{+ 1.3\%}_{-1.4 \%} \pm 2.4 \% $ \\
14 & 125.09 & $0.196^{+1.3 \%}_{-1.4 \%} \pm 2.4 \% $ &$0.112^{+ 1.3\%}_{-1.3 \%} \pm 2.5 \%$ &  $0.325^{+ 1.3\%}_{-1.4 \%} \pm 2.4 \%$ & $0.496^{+ 1.3\%}_{-1.4 \%} \pm 2.4 \% $ \\
14 & 125.38 &  $ 0.195^{+1.3 \%}_{-1.4\%} \pm 2.4 \% $ &$0.111^{+ 1.3\%}_{-1.3 \%} \pm 2.5 \%$ &  $0.322^{+ 1.3\%}_{-1.4 \%} \pm 2.4 \%$ & $0.493^{+ 1.3\%}_{-1.4 \%} \pm 2.4 \% $\\
\end{tabular}}
\caption{Cross section values for $W^-hh$ production including scale uncertainties and PDF uncertainties in percent in this order at NNLO QCD in the SM ($\kappa_\lambda = 1$) and for three anomalous values of the trilinear coupling $\kappa_\lambda=\{0,2,3\}$.
\label{tab:Wmhh} }
\end{table}

\subsection{$Zhh$}
A peculiarity of $Zhh$ production is that, at NNLO QCD, gluon-initiated contributions arising from heavy-fermion triangle, box, and pentagon diagrams must be taken into account. In the computation of the triangle, box and pentagon diagrams a bottom mass of $m_b=4.9$~GeV is set. Those contributions have been computed in \cite{Baglio:2012np} on which the numbers presented here are based. In \cite{Agrawal:2017cbs} differential distributions and anomalous couplings of $gg\to Zhh$ have been considered.
The values of the cross section for various centre-of-mass energies and Higgs boson masses can be found in Table~\ref{tab:Zhh}. The scale uncertainties are significantly higher than for $W^{\pm}hh$ due to the additional $gg\to Zhh$ contributions that as typical for gluon fusion processes exhibit large scale uncertainties at LO.\footnote{Similarly to $gg\to Zh$ one expects also a large top mass renormalisation uncertainty \cite{Chen:2022rua, Degrassi:2022mro, CampilloAveleira:2025rbh} not computed here.}
\begin{table}
\renewcommand{\arraystretch}{1.5}
\small
\centerline{
\begin{tabular}{c| c |c | c | c| c }
$\sqrt{s}$ [TeV] & $m_h$ [GeV] & $\sigma $ [fb] SM & $\sigma $ [fb] $\kappa_{\lambda}=0$ & $\sigma $ [fb] $\kappa_{\lambda}=2$  & $\sigma $ [fb] $\kappa_{\lambda}=3$  \\ \hline
13 & 125 & $0.366^{+ 3.3\%}_{-2.6 \%} \pm 2.0 \%$ & $0.232^{+ 5.1\%}_{-4.0 \%} \pm 1.9\%$ & $0.561^{+ 2.2\%}_{-1.8 \%} \pm 2.0\%$ & $0.819^{+ 1.5\%}_{-1.3 \%} \pm 2.0 \%$ \\
13 & 125.09 & $0.365^{ +3.3\%}_{-2.6 \%} \pm 2.0\%$ & $0.232^{+ 5.1\%}_{-4.0 \%} \pm 1.9\%$ & $0.560^{+ 2.2\%}_{-1.8 \%} \pm 2.0\%$ & $0.818^{+ 1.5\%}_{-1.3 \%} \pm 2.0\%$  \\
13 & 125.38 & $0.362 ^{+3.3 \%}_{-2.6 \%} \pm 2.0 \%$ & $0.230^{+ 5.1\%}_{-4.0 \%} \pm 1.9\%$ & $0.556^{+ 2.2\%}_{-1.8 \%} \pm 2.0\%$ & $0.813^{+ 1.5\%}_{-1.3 \%} \pm 2.0 \%$  \\ \midrule
13.6 & 125 & $0.396 ^{+3.4 \%}_{-2.7 \%}  \pm 1.9 \%$ & $0.253 ^{+5.3 \%}_{-4.1 \%}  \pm 1.9 \%$ & $0.607 ^{+2.2 \%}_{-1.8 \%}  \pm 2.0 \%$ & $0.885 ^{+1.6 \%}_{-1.3 \%}  \pm 2.0 \%$\\
13.6 & 125.09 & $ 0.396^{+3.4\%}_{-2.7\%}  \pm 1.9 \% $ & $0.252 ^{+5.3 \%}_{-4.1 \%}  \pm 1.9 \%$ & $0.606 ^{+2.2 \%}_{-1.8 \%}  \pm 2.0 \%$ & $0.883 ^{+1.6 \%}_{-1.3 \%}  \pm 2.0 \%$ \\
13.6 & 125.38 & $0.393^{+3.4 \%}_{-2.7 \%}  \pm 1.9 \% $ & $0.250 ^{+5.3 \%}_{-4.1 \%}  \pm 1.9 \%$ & $0.602 ^{+2.2 \%}_{-1.8 \%}  \pm 2.0 \%$ & $0.878 ^{+1.6 \%}_{-1.3 \%}  \pm 2.0 \%$ \\ \midrule
14 & 125 &  $0.417^{+3.4\%}_{-2.7 \%} \pm 1.9 \% $ &   $0.267^{+5.3\%}_{-4.1 \%} \pm 1.8 \% $ & $0.638^{+2.3\%}_{-1.8 \%} \pm 1.9 \% $ &  $0.929^{+1.6\%}_{-1.3 \%} \pm 2.0 \% $  \\
14 & 125.09 & $0.416^{+3.4 \%}_{-2.7 \%} \pm 1.9 \% $ & $0.266^{+5.3\%}_{-4.1 \%} \pm 1.8 \% $ & $0.637^{+2.3\%}_{-1.8 \%} \pm 1.9 \% $ &  $0.927^{+1.6\%}_{-1.3 \%} \pm 2.0 \% $ \\
14 & 125.38 &  $ 0.413^{+3.4 \%}_{-2.7\%} \pm 1.9 \% $ & $0.264^{+5.3\%}_{-4.1 \%} \pm 1.8 \% $ & $0.632^{+2.3\%}_{-1.8 \%} \pm 1.9 \% $ &  $0.922^{+1.6\%}_{-1.3 \%} \pm 2.0 \% $\\
\end{tabular}}
\caption{Cross section values for $Zhh$ production including scale uncertainties and PDF uncertainties in percent in this order at NNLO QCD in the SM ($\kappa_\lambda = 1$) and for three anomalous values of the trilinear coupling $\kappa_\lambda=\{0,2,3\}$. \label{tab:Zhh} }
\end{table}

\section{Conclusion}
In this note we have presented updated inclusive cross sections for Higgs boson pair production for both Higgs Strahlung and VBF. The cross sections are presented at the highest perturbative orders, and at energies of relevance to both the LHC and HL-LHC. The numbers presented here should serve as a valuable reference to the experiments.
\section*{Acknowledgements}
This work was done on behalf of the LHC Higgs Working Group. RG received funding by the INFN Iniziativa Specifica APINE and by the University of Padua under the 2023 STARS Grants@Unipd programme (Acronym and title of the project: HiggsPairs – Precise Theoretical Predictions for Higgs pair production at the LHC). This work was also partially supported by the Italian MUR Departments of Excellence grant 2023-2027 “Quantum Frontiers”. AK acknowledges funding from a Royal Society
Research Professorship (grant RP$\backslash$R$\backslash$231001), and wishes to thank the CERN Theory Department for hospitality while this work was carried out.
MP acknowledges support by the state of Baden-Württemberg
through bwHPC and the German Research Foundation (DFG) through grant No.\ INST 39/963-1 FUGG (bwForCluster NEMO). We acknowledge support by the COST Action COMETA CA22130. 
\bibliography{bibliography}
\end{document}